\documentclass[reprint,amsmath,amssymb,aip,jcp,floatfix]{revtex4-1}

\usepackage{amssymb}
\usepackage{amsmath}
\usepackage{graphicx}
\usepackage{amsbsy}
\usepackage{color}
\usepackage{bm}
\usepackage{dcolumn}

\newcommand{\bq}{\begin{eqnarray}}
\newcommand{\eq}{\end{eqnarray}}
\newcommand{\bqn}{\begin{eqnarray*}}
\newcommand{\eqn}{\end{eqnarray*}}

\newcommand{\rr}{\mathbf{r}}

\newcommand{\hs}{\text{HS}}
\newcommand{\cs}{\text{CS}}
\newcommand\beq{\begin{equation}}
\newcommand\eeq{\end{equation}}
\newcommand\beqa{\begin{eqnarray}}
\newcommand\eeqa{\end{eqnarray}}
\newcommand{\nn}{\nonumber\\}

\begin{document}
\title{Phase {diagrams} of  Janus fluids with up-down constrained orientations}

\author{Riccardo Fantoni}
\email{rfantoni@ts.infn.it}

\author{Achille Giacometti}
\email{achille.giacometti@unive.it}
\affiliation{Dipartimento di Scienze dei Materiali e Nanosistemi, Universit\`a Ca' Foscari Venezia,
Calle Larga S. Marta DD2137, I-30123 Venezia, Italy}

\author{Miguel \'Angel G. Maestre}
\email{maestre@unex.es}

\author{Andr\'es Santos}
\email{andres@unex.es}
\homepage{http://www.unex.es/eweb/fisteor/andres}
\affiliation{Departamento de F\'{\i}sica, Universidad de Extremadura,
E-06071 Badajoz, Spain}

\date{\today}

\begin{abstract}
A class of binary mixtures of Janus fluids formed by colloidal spheres with the hydrophobic hemispheres constrained to
point either up or down are studied by means of Gibbs ensemble Monte Carlo simulations
and  simple analytical approximations.
These fluids can be experimentally realized by
the application of an external static electrical field. {The gas-liquid and demixing phase transitions in five} specific models with different patch-patch affinities are analyzed. It is found that  a gas-liquid transition is {present in all the models}, even if only one of the four possible patch-patch interactions is attractive. {Moreover, provided the attraction between like particles is stronger than between unlike particles,} the system demixes into two subsystems with different composition at sufficiently low temperatures and high densities.
\end{abstract}

\maketitle

\section{Introduction\label{sec1}}
Engineering new materials through direct self-assembly processes has recently become
a new concrete possibility due to the remarkable developments in the synthesis of patchy colloids with different shapes and
functionalities. Nowadays, both the synthesis and the aggregation process of patchy colloids can be experimentally controlled
with a precision and reliability that were not possible until a few years ago.\cite{GS07,WM08,HCLG08,PK10,WM13}

Within the general class of patchy colloids, a particularly
interesting case is provided by the so-called Janus
fluid, where the surface
of the colloidal particle is evenly partitioned between the hydrophobic
and the hydrophilic moieties, so that attraction between two spheres is  possible only if both hydrophobic  patches
are facing one another.\cite{BBL11} Several experimental and theoretical studies have illustrated the remarkable properties  of this paradigmatic
case.\cite{JG12,F13}
%

The behavior  of patchy particles under external fields has received recent attention.\cite{GPKV10,G11} By applying an external electrical or magnetic field, appropriately
synthesized dipolar Janus particles may be made to align orientationally, so
as to expose their functionally active hemisphere either all up or all
down (See Ref.\ \onlinecite{G11},  Secs.\ 1.4.3.1 and 1.4.3.2, and references
therein). By mixing the two species one could have in the laboratory a
binary mixture of Janus particles where the functionally active patch
points in opposite directions for each species.

While theoretical studies have been keeping up with, and sometimes even anticipated, experimental
developments, the complexities of the anisotropic interactions in patchy colloids
have mainly restricted these investigations to numerical simulations, which have revealed interesting specificities in the
corresponding phase diagrams.

Motivated by the above scenario, we have recently introduced a simplified binary-mixture model of a fluid of Janus spheres (interacting via the anisotropic Kern--Frenkel potential),\cite{KF03} where the hydrophobic patches
on each sphere  could point only either up (species 1) or down (species 2).\cite{MFGS13}
This orientational restriction, which is reminiscent of Zwanzig's model for liquid crystals, clearly simplifies the theoretical description while
still distilling out the main features of the original Janus model.

\begin{figure}
\begin{center}
\includegraphics[width=7cm]{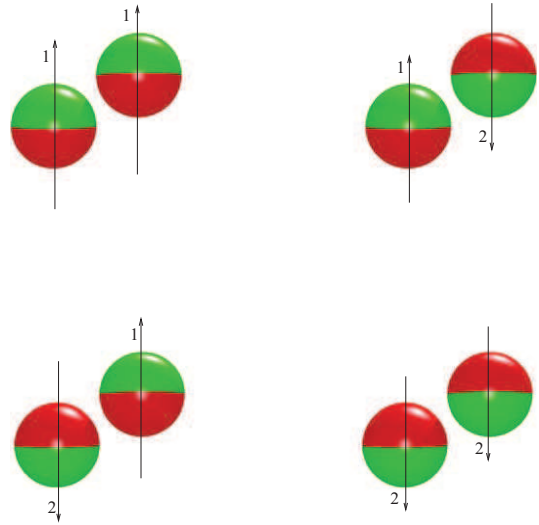}
\end{center}
\caption{Sketch of a binary-mixture Janus fluid with up-down constrained orientations. The energy scales of the attractive interactions are (from left to right and from top to bottom) $\epsilon_{11}$, $\epsilon_{12}$, $\epsilon_{21}$, and $\epsilon_{22}=\epsilon_{11}$, respectively. {Here we have adopted the convention that $\epsilon_{ij}$ is the energy  scale when a particle of species $i$ is ``below'' a particle of species $j$.}}
\label{fig:fig1}
\end{figure}

In the present paper, we generalize the above Janus fluid model by assuming arbitrary values for {the} energy scales $\epsilon_{ij}$ {of the attractive interactions} associated with the four possible pair configurations (see Fig.\ \ref{fig:fig1}), which allows for a free tuning of the strength of the patch-patch attraction. In some cases this can effectively mimic the reduction of the coverage in the original Kern--Frenkel model.
{Note that, in Fig.\ \ref{fig:fig1}, $\epsilon_{ij}$ is the energy associated with the (attractive) interaction between a particle of species $i$ (at the left) and a particle of species $j$ (at the right) when the former is below the latter,
with the arrow always indicating the hydrophobic (i.e. attractive) patch. The original Kern--Frenkel model then corresponds to $\epsilon_{12}>0$ and $\epsilon_{11}=\epsilon_{22}
=\epsilon_{21}=0$, whereas  the full coverage limit is equivalent to $\epsilon_{11}=\epsilon_{22}=\epsilon_{12}=\epsilon_{21}>0$. On the other hand, the effect of
reducing the coverage from the full to the Janus limit, can be effectively mimicked by fixing $\epsilon_{12}>0$ and progressively decreasing $\epsilon_{21}$ and $\epsilon_{11}=\epsilon_{22}$.}
Moreover, the class of models depicted in Fig.\ \ref{fig:fig1} allows for an interpretation more general and flexible than the hydrophobic-hydrophilic one. For instance, one may assume that attraction is only possible when patches of \emph{different} type are facing one another (i.e., $\epsilon_{11}=\epsilon_{22}>0$ and $\epsilon_{12}=\epsilon_{21}=0$).
As shown below, this will provide a rich scenario of intermediate cases with a number of  interesting features in the phase diagram of both the gas-liquid and the
demixing transitions.

{We emphasize the fact that in the simulation part of the present study we will always assume ``global'' equimolarity, that is, the combined number of particles of species $1$ ($N_1$) is
always equal to the combined number of particles of species $2$ ($N_2$), so that $N_1=N_2=N/2$, where $N$ is the total number of particles. On the other hand, the equimolarity condition is not imposed on each coexisting phase.}

The organization of the paper is as follows. The class of models is briefly described in Sec.\ \ref{sec2}. Next, in Sec.\ \ref{sec3} we present our Gibbs ensemble Monte Carlo (GEMC) results for the gas-liquid and demixing transitions. The complementary theoretical approach is presented in Sec.\ \ref{sec4}. The paper is closed with some concluding remarks in Sec.\ \ref{sec5}.

\section{Description of the models\label{sec2}}
\begin{figure}
\begin{center}
  \includegraphics[width=8cm]{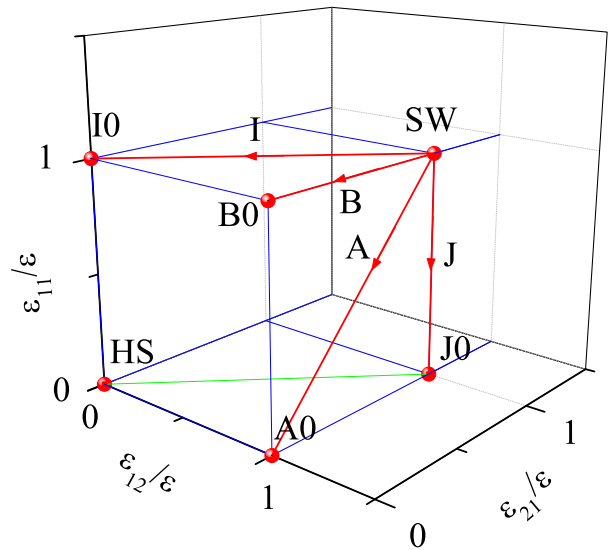}
\end{center}
\caption{Parameter space of the class of Janus models defined in the paper.}
\label{fig:fig2}
\end{figure}

In our class of binary-mixture Janus models, particles of species 1 (with a mole fraction $x_1$) and 2 (with a mole fraction $x_2=1-x_1$) are dressed with two up-down hemispheres with different attraction properties, as sketched in Fig.\ \ref{fig:fig1}. The pair potential between a particle of species $i$ at $\rr_1$ and a particle  of species $j$ at $\rr_2$ is
\beq
\phi_{ij}(\rr_{12})=\varphi_{ij}(r_{12})\Theta(z_{12})+
\varphi_{ji}(r_{12})\Theta(-z_{12}),
\eeq
where $\Theta(z)$ is the Heaviside step function, $\rr_{12}=\rr_2-\rr_1$, $z_{12}=z_2-z_1$, and
\beq
\varphi_{ij}(r)=\begin{cases}
\infty,       &   0\le r< \sigma ,\\
-\epsilon_{ij},&   \sigma\le r< \sigma +\Delta,  \\
0,            &   \sigma +\Delta\le r,
\end{cases}
\eeq
is a standard square-well (SW) potential of diameter $\sigma$, width $\Delta$, and energy depth $\epsilon_{ij}$, except that, in general, $\epsilon_{12}\neq\epsilon_{21}$.
By symmetry, {one must have} $\epsilon_{22}=\epsilon_{11}$ {(see Fig.\ \ref{fig:fig1})}, so that (for given values of $\sigma$ and $\Delta$) the space parameter of the interaction potential becomes three-dimensional, as displayed in Fig.\ \ref{fig:fig2}. Except in the case of  the hard-sphere (HS) model ($\epsilon_{ij}=0$), one can freely choose one of the non-zero $\epsilon_{ij}$ to fix the energy scale. Thus, we call $\epsilon=\max_{i,j}\{\epsilon_{ij}\}$ and use the three independent ratios $\epsilon_{ij}/\epsilon$ as axes in Fig.\ \ref{fig:fig2}. The  model represented by the coordinates $(1,1,1)$ is the fully isotropic SW fluid, where species 1 and 2 become indistinguishable. Next, without loss of generality, we choose $\epsilon_{12}\geq \epsilon_{21}$.
With those  criteria, all possible models of the class lie either inside the triangle SW-I0-B0-SW or inside the square SW-B0-A0-J0-SW.
{One could argue that any point \emph{inside} the cube displayed in Fig.\ \ref{fig:fig2} may represent a distinct model, but this is not so. First, the choice $\epsilon=\max_{i,j}\{\epsilon_{ij}\}$ restricts the models to those lying on one of the three faces $\epsilon_{11}/\epsilon=1$, $\epsilon_{12}/\epsilon=1$, or $\epsilon_{21}/\epsilon=1$. Second, the choice $\epsilon_{12}\geq \epsilon_{21}$ reduces the face  $\epsilon_{21}/\epsilon=1$ to the line SW-J0 and the face $\epsilon_{11}/\epsilon=1$ to the half-face SW-I0-B0-SW.}
The vertices {SW, I0, B0, A0, and J0} define the five distinguished models we will specifically study. Those models, together with the HS one, are summarized in Table \ref{tab:models}.

%

\begin{table}
\caption{{Definition of the models.}}
\label{tab:models}
\begin{ruledtabular}
\begin{tabular}{ccccc}
Model &$\epsilon_{11}$  &$\epsilon_{12}$  &$\epsilon_{21}$  &$\epsilon_{22}$ \\
\hline
HS    &$0$              &$0$              &$0$              &$0$             \\
A0    &$0$              &$\epsilon$       &$0$              &$0$             \\
I0    &$\epsilon$       &$0$              &$0$              &$\epsilon$      \\
J0    &$0$              &$\epsilon$       &$\epsilon$       &$0$             \\
B0    &$\epsilon$       &$\epsilon$       &$0$              &$\epsilon$      \\
SW   &$\epsilon$       &$\epsilon$       &$\epsilon$       &$\epsilon$      \\
\end{tabular}
\end{ruledtabular}
\end{table}

{The rationale behind our nomenclature for the models goes as follows. Models with $\epsilon_{12}=\epsilon_{21}$ are isotropic and so we use the letter I to denote the isotropic models with $0\leq \epsilon_{12}/\epsilon=\epsilon_{21}/\epsilon\leq 1$ and $\epsilon_{11}/\epsilon=1$. Apart from them,  the only additional isotropic models are those with $\epsilon_{12}/\epsilon=\epsilon_{21}/\epsilon=1$ and $0\leq\epsilon_{11}/\epsilon\leq 1$, and we denote them with the {letter (J) next to I}. All the remaining models are anisotropic (i.e., $\epsilon_{12}\neq\epsilon_{21}$). Out of them, we use the letter A to denote the particular subclass of  anisotropic models ($0\leq \epsilon_{11}/\epsilon=\epsilon_{21}/\epsilon\leq 1$ and $\epsilon_{12}/\epsilon=1$) which can be viewed as the anisotropic counterpart of the isotropic subclass I. Analogously, we employ the {letter (B) next to A} to refer to the anisotropic counterpart ($\epsilon_{11}/\epsilon=\epsilon_{12}/\epsilon=1$ and $0\leq\epsilon_{21}/\epsilon\leq 1$) of the isotropic models J. Finally, the number 0 is used to emphasize that the corresponding models are the extreme cases of the subclasses I, J, A, and B, respectively.}

Model A0  is the one more directly related to the original Kern--Frenkel potential and was the one analyzed in Ref.\ \onlinecite{MFGS13}. Also related to that potential is model B0, where only the interaction between the two hydrophilic patches is purely repulsive. On the other hand, {in models I0 and J0} (where $\epsilon_{12}=\epsilon_{21}$) the interaction becomes isotropic and the Janus character of the model is blurred.
In model I0 the fluid reduces to
a binary mixture with attractive interactions between like components and HS repulsions between unlike ones. This model was previously studied by Zaccarelli et al.\cite{ZFTSD00} using integral equation techniques.
In the complementary model J0 attraction exists only between unlike particles. The points A0, B0, I0, and J0 can be reached from the one-component SW fluid along models represented by the lines A, B, I, and J, respectively. Of course, other intermediate models are possible inside the triangle SW-I0-B0-SW or {inside} the square SW-B0-A0-J0-SW.

{In addition to the energy parameters $\epsilon_{ij}$, the number density $\rho$, and the temperature $T$, each particular system is specified by the mixture composition (i.e., the mole fraction $x_1$). In fact, in Ref.\ \onlinecite{MFGS13} the thermodynamic and structural properties of model A0 were studied both under equimolar and non-equimolar conditions.}

\section{Gibbs ensemble Monte Carlo simulations}
\label{sec3}

In this paper, we use GEMC techniques\cite{P87,SSF89,SF89} to study the gas-liquid condensation process
of  models SW, A0, B0, I0, and J0 {and the demixing transition of models I0 and B0}.
We have chosen the width of the active attractive  patch as in the
experiment of Hong et al.\cite{HCLG08} ($\Delta/\sigma=0.05$).
Given the very small width of the attractive wells, we expect the liquid phase to be  metastable
with respect to the corresponding solid one.\cite{LGK05,MF03,VPSDS13}
Reduced densities  $\rho^*=\rho\sigma^3$ and temperatures $T^*=k_BT/\epsilon$ will be employed throughout.

\subsection{Technical details}
The GEMC  method  is  widely adopted as a
standard method for calculating phase equilibria from molecular
simulations. According to this method, the simulation is performed in
two boxes (I and II) containing the coexisting phases. Equilibration in each
phase is guaranteed by moving particles. Equality of pressures is
satisfied in a statistical sense by expanding the volume of one of the boxes and contracting the volume of the other one, keeping the total volume constant. Chemical potentials are
equalized by transferring particles from one box to the other one.

In the GEMC run we have on each step a probability $a_p/(a_p+a_v+a_s)$,  $a_v/(a_p+a_v+a_s)$, and  $a_s/(a_p+a_v+a_s)$
for a particle random displacement,  a volume
change, and  a particle swap move between both boxes, respectively. {We generally chose the relative weights $a_p=1$, $a_v=1/10$, and
$a_s=20$.} {To preserve the up-down fixed patch orientation, rotation of particles was not allowed}. The maximum particle displacement was kept equal to $10^{-3}L^{(\gamma)}$ where
$L^{(\gamma)}$ is the side of the (cubic) box $\gamma=$I, II. Regarding the volume
changes, following Ref.\ \onlinecite{FS02} we performed a random walk
in $\ln(V^\text{(I)}/V^\text{(II)})$, with $V^{(\gamma)}$ the volume of the box $\gamma$, choosing a
maximum volume displacement of $1\%$.
The volume move is
computationally the most expensive one. {This is because, after each volume move, it is necessary, in order to determine the next acceptance probability, to perform a full potential energy calculation since \emph{all} the
particle coordinates are rescaled by the factor associated with the
enlargement or reduction of the boxes. However, this is not necessary for the
other two moves since in those cases only the coordinates of a single particle change.}

Both in the condensation and   in the demixing problems, the Monte Carlo swap move consisted in moving a particle selected randomly in one box into the other box, so that the number of particles of each species in both boxes ($N_1^\text{(I)}$, $N_2^\text{(I)}$, $N_1^\text{(II)}$, and $N_2^\text{(II)}$) were fluctuating quantities. The only constraint was that the {\emph{total}} number of particles  was the same for both species, i.e., $N_1\equiv N_1^\text{(I)}+N_1^\text{(II)}=N_2^\text{(I)}+N_2^\text{(II)}\equiv N_2=N/2$. In the condensation problem we fixed the \emph{global} density $\rho=N/(V^\text{(I)}+V^\text{(II)})$ ({in all the cases we took $\rho^*=0.3$, a value slightly below the expected critical density}) and then varied the temperature $T$ (below the critical temperature). The measured output quantities where the partial densities $\rho^\text{(I)}=N^\text{(I)}/V^\text{(I)}$ and $\rho^\text{(II)}=N^\text{(II)}/V^\text{(II)}$, where $N^{(\gamma)}=N_1^{(\gamma)}+N_2^{(\gamma)}$ is the total number of particles in box $\gamma=${I, II}. Note that $(\rho^\text{(II)}-\rho)/(\rho-\rho^\text{(I)})=V^\text{(I)}/V^\text{(II)}$.   In contrast, in the demixing problem we fixed $T$ (above the critical temperature) and varied  $\rho$, the output observables being the local mole fractions $x_1^\text{(I)}=N_1^\text{(I)}/N^\text{(I)}$ and $x_1^\text{(II)}=N_1^\text{(II)}/N^\text{(II)}$. In this case, the lever rule is $(x_1^\text{(II)}-\frac{1}{2})/(\frac{1}{2}-x_1^\text{(I)})=N^\text{(I)}/N^\text{(II)}$.

{The total number of  particles {of each species was $N_1=N_2=250$}, what was checked to be sufficient for our purposes.}  We used $50\text{--}100\times 10^6$
 MC steps for the equilibration (longer near the critical point) and $100\text{--}200\times 10^6$ MC steps for the production.\cite{note_13_06}

\subsection{Gas-liquid coexistence}

\begin{figure}
\begin{center}
  \includegraphics[width=8.5cm]{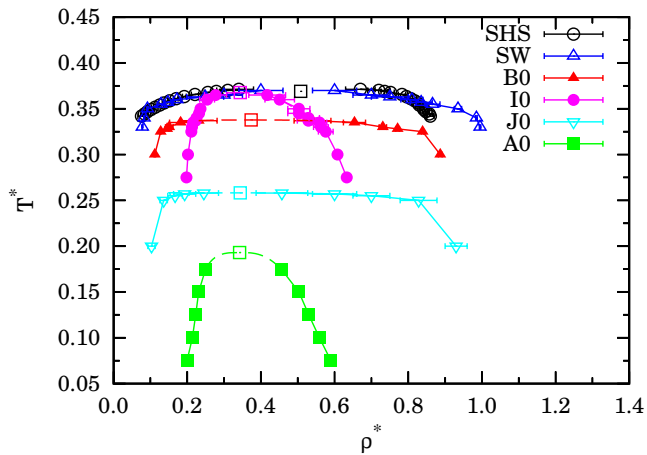}
\end{center}
\caption{Gas-liquid binodals for  models SW, B0, I0,  J0, and A0. The points indicated as
SHS in the legend are grand canonical MC (GCMC) results taken from Ref.\ \onlinecite{MF04}, where the actual one-component SHS model was studied. The remaining results  are
those obtained in this work from GEMC simulations. In each case, the solid line is  a guide to the eye, while the dashed line is the result  of the extrapolation to the critical point, which is represented by a square. }
\label{fig:fig3}
\end{figure}

\begin{table*}
\caption{{Gas-liquid coexistence properties for  models A0, B0, I0, and J0, as obtained from our GEMC simulations.
$T^*$ is the reduced temperature, $\rho^*_\gamma$ is the reduced density  of the gas  ($\gamma=g$) and  liquid ($\gamma=l$) phases,  $N^{(g)}$ is the average number of particles in the gas box, and $U_\text{ex}^{(\gamma)}/N^{(\gamma)}$ is the  excess internal energy {per particle in box $\gamma$}.}}
\label{tab:gemc}
\begin{ruledtabular}
\begin{tabular}{cdddddd}
Model&\multicolumn{1}{c}{$\;\;\;\;\;T^*$} &\multicolumn{1}{c}{$\;\;\;\;\;\;\;\rho^*_g$} & \multicolumn{1}{c}{$\;\;\;\;\;\;\;\rho^*_l$}&\multicolumn{1}{c}{$\;\;\;\;\;\;\;N^{(g)}/N$} & \multicolumn{1}{r}{$\;\;\;\;-U_\text{ex}^{(g)}/\epsilon N^{(g)}$} & \multicolumn{1}{r}{$\;\;\;-U_\text{ex}^{(l)}/\epsilon N^{(l)}$}\\
\hline
A0 & 0.075 &0.1994(6) & 0.590(1)& 0.493(2) & 1.69(1) & 1.796(7)  \\
&  0.1 &0.214(2) & 0.559(5)& 0.535(4) & 1.785(4) & 1.780(8)  \\
&  0.125 & 0.223(1) & 0.530(6)& 0.556(3) & 1.63(9) & 1.71(5)  \\
&  0.15 &0.231(1) & 0.503(4) & 0.574(4) & 1.60(1) & 1.78(1) \\
&  0.175&0.250(2) & 0.455(8)  & 0.630(6) & 1.42(1) & 1.632(9) \\
 &&&&&&\\
B0&  0.3 &0.112(2) & 0.887(5)& 0.284(5) & 1.6(1) & 3.27(1)  \\
&  0.325&0.128(1) & 0.839(3) & 0.324(3) & 0.761(1) & 3.239(7)  \\
&  0.328 &0.145(5) & 0.771(5) & 0.363(9) & 0.88(2) & 2.99(1)  \\
&  0.33 &0.15(1) & 0.73(1)& 0.380(1) & 0.95(1) & 3.016(9)  \\
&  0.335 &0.18(3) & 0.65(3)  & 0.45(1) & 1.0(7) & 2.83(2) \\
& 0.337  &0.23(5) & 0.54(5) & 0.59(1) & 1.273(4) & 2.36(4)\\
 &&&&&&\\
I0&  0.3&0.202(3) & 0.61(1)  & 0.5146(7) & 2.48(6) & 3.04(1) \\
&  0.325&0.211(5) & 0.58(2) & 0.5371(6) & 1.76(4) & 2.765(8)  \\
&  0.35 &0.24(1) & 0.50(3) & 0.612(3) & 1.24(3) & 2.30(1) \\
&  0.36&0.25(2) & 0.45(4) & 0.657(5) & 1.01(1) & 1.85(5)  \\
&  0.365  &0.28(3) & 0.42(5)& 0.71(1) & 0.96(2) & 1.6(1) \\
 &&&&&&\\
J0& 0.2  & 0.10(1) & 0.93(3)& 0.249(5) & 1.67(2) & 2.48(3) \\
&  0.25 &0.14(1) & 0.83(5) & 0.34(1) & 0.82(2) & 2.25(3) \\
&  0.255&0.17(2) & 0.70(5) & 0.433(9) & 0.90(2) & 1.99(2)  \\
&  0.257&0.19(3) & 0.60(6) & 0.62(6) & 1.10(7) & 1.5(2)  \\
\end{tabular}
\end{ruledtabular}
\end{table*}

\begin{table}
\caption{{Mole fractions in the gas and liquid boxes in model I0  at different temperatures and with a global density $\rho^*=0.3$. For the gas and liquid densities, see Table \ref{tab:gemc}. Because of the symmetry under label exchange $1\leftrightarrow 2$, we have adopted the criterion $x_1^{(g)}\leq x_2^{(g)}$ without loss of generality.}}
\label{table2b}
\begin{ruledtabular}
\begin{tabular}{ddd}
\multicolumn{1}{c}{$\;\;\;\;\;T^*$} &\multicolumn{1}{c}{$\;\;\;\;\;\;\;x_1^{(g)}$} &\multicolumn{1}{c}{$\;\;\;\;\;\;\;x_1^{(l)}$}\\
\hline
0.3  &                  0.03(1) &                   0.992(6)\\
0.325    &           0.09(2)  &                    0.98(1) \\
0.35      &          0.18(3)  &                     0.955(15)\\
0.36       &         0.26(3)   &                   0.93(3)\\
0.365       &       0.34(3)     &                0.89(4) \\
\end{tabular}
\end{ruledtabular}
\end{table}

Results for the gas-liquid transition  are depicted in Fig.\ \ref{fig:fig3} in the temperature-density plane. {Some representative numerical values} for models A0, B0, I0, and J0 are tabulated in Table \ref{tab:gemc}.  In this case, one of the
two simulation boxes (I=$g$) contains the gas phase and the
other one  (II=$l$) contains the liquid phase. {Since $\rho_g<\rho<\rho_l$, the choice of the global density $\rho$ establishes a natural bound as to how close  to the critical point the measured binodal curve can be. In fact, $N^{(g)}\to 0$ if $\rho_l\to \rho$, while $N^{(g)}\to N$ if $\rho_g\to \rho$. As is apparent from the values of $N^{(g)}/N$ in Table \ref{tab:gemc}, the latter scenario seems to take place in our case $\rho^*=0.3$.}

{Although not strictly enforced, we observed that $N_1^{(g)}\simeq N_2^{(g)}$ and $N_1^{(l)}\simeq N_2^{(l)}$ (so  both boxes were practically equimolar) in models A0, B0, and J0. On the other hand, in the case of model I0 the final equilibrium state was non-equimolar (despite the fact that, as said before, $N_1=N_2$ globally), the low-density box having a more disparate composition than the high-density box. The mole fraction values are shown in Table \ref{table2b}. Thus, in contrast to models A0, B0, and J0, the GEMC simulations at fixed temperature and global density $\rho^*=0.3$  spontaneously drove the system I0 into two coexisting boxes differing both in density and composition. This \emph{spontaneous demixing} phenomenon means that in model I0 the equimolar binodal curve must be metastable with respect to demixing and so it was not observed in our simulations. It is important to remark that, while the equimolar binodal must be robust with respect to changes in the global density $\rho$ (except for the bound $\rho_g<\rho<\rho_l$ mentioned above), the non-equimolar binodal depends on the value of $\rho$.}

In addition to cases SW, B0, I0, J0, and A0,  we have also included in Fig.\ \ref{fig:fig3}, for completeness, numerical results
obtained by Miller and Frenkel\cite{MF04} on the one-component Baxter's sticky-hard-sphere (SHS) model.\cite{B68} As expected, they agree quite well with our short-range SW results, the only qualitative difference being a liquid branch at slightly larger densities.

In order to determine the critical point $(T_c^*,\rho_c^*)$ we
empirically extrapolated the GEMC binodals using the law of
rectilinear ``diameters'',\cite{SLS78}
$\frac{1}{2}\left(\rho_g^{*}+\rho_l^{*}\right)=\rho_c^*+A|T^*-T_c^*|$,
and the Wegner expansion\cite{W72,SLS78} for the
width of the coexistence curve,
$\rho_l^*-\rho_g^*=B|T^*-T_c^*|^{\beta_I}$.  The critical coordinates  $(T_c^*,\rho_c^*)$ and the coefficients
$A$ and $B$ are taken as fitting parameters. {The {four} points corresponding to the  two highest temperatures  were used for the extrapolation in each case}. We remark that our data do not extend
sufficiently close to the critical region to allow for quantitative
estimates of critical exponents and non-universal quantities. However, assuming that the models  belong to the three-dimensional Ising
universality class,  we chose $\beta_I=0.325$. The numerical values obtained by this extrapolation procedure will be  presented in Table \ref{tab:crit} below.

The decrease in the critical temperatures and densities in going from the one-component SW fluid to model B0
and then to model A0  is strongly reminiscent of an analogous trend present in the unconstrained one-patch
Kern--Frenkel model upon decrease of the coverage. \cite{SGP10}

\begin{figure}
\begin{center}
\includegraphics[width=9cm]{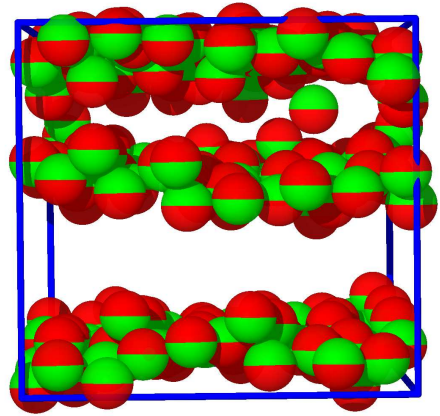}
\end{center}
\caption{{Snapshot of the liquid-phase box in model A0 at  $T^*=0.15$.}}
\label{fig:snap}
\end{figure}

It is interesting to remark that,
even though the influence of attraction in model A0 is strongly inhibited by the up-down constrained orientation {($\epsilon_{ij}=\epsilon\delta_{i1}\delta_{j2}$)},
this model exhibits a gas-liquid transition. {This surprising result was preliminarily supported by canonical $NVT$ MC simulations in Ref.\ \onlinecite{MFGS13}, but now it is confirmed by the new and more appropriate GEMC simulations presented in this paper. Given the patch geometry and interactions in model A0, one might expect the formation of a lamellar-like liquid phase  (approximately) made of {alternating layers (up-down-up-down-$\cdots$) of particles with the same orientation}. This scenario is confirmed by snapshots of the liquid-phase box, as illustrated by Fig.\ \ref{fig:snap}.}

The Kern--Frenkel analogy is not applicable to the isotropic models I0 and J0.  Model J0 presents a critical point  intermediate between those of models B0 and A0, as expected.  However, while the decrease in the total average attractive
strength is certainly one of the main mechanisms dictating the location of the gas-liquid coexistence curves, it {cannot be the only discriminating factor, as shown by the results for the isotropic model I0, where the critical temperature is higher and the binodal curve is narrower than that corresponding to the anisotropic model B0. This may be due to the fact that, as said before, the binodal curve in model I0 is not equimolar and this lack of equimolarity is expected to extend to the critical point, as can be guessed from the trends observed in Table \ref{table2b}. In other words, two demixed phases can be made to coexist at a higher temperature and with a smaller density difference than two mixed phases.}


\begin{figure}
\begin{center}
\includegraphics[width=8cm]{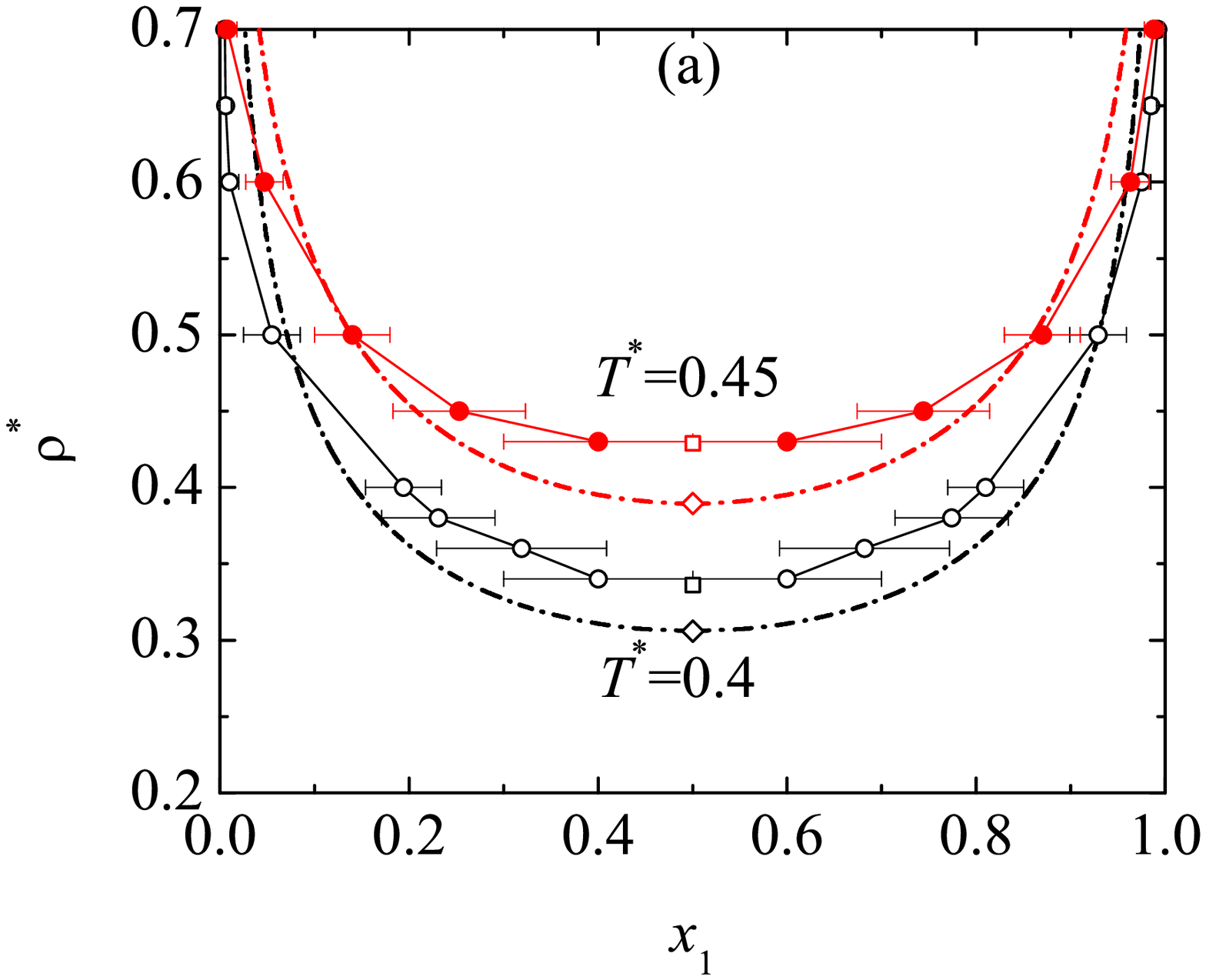}\\
\includegraphics[width=8cm]{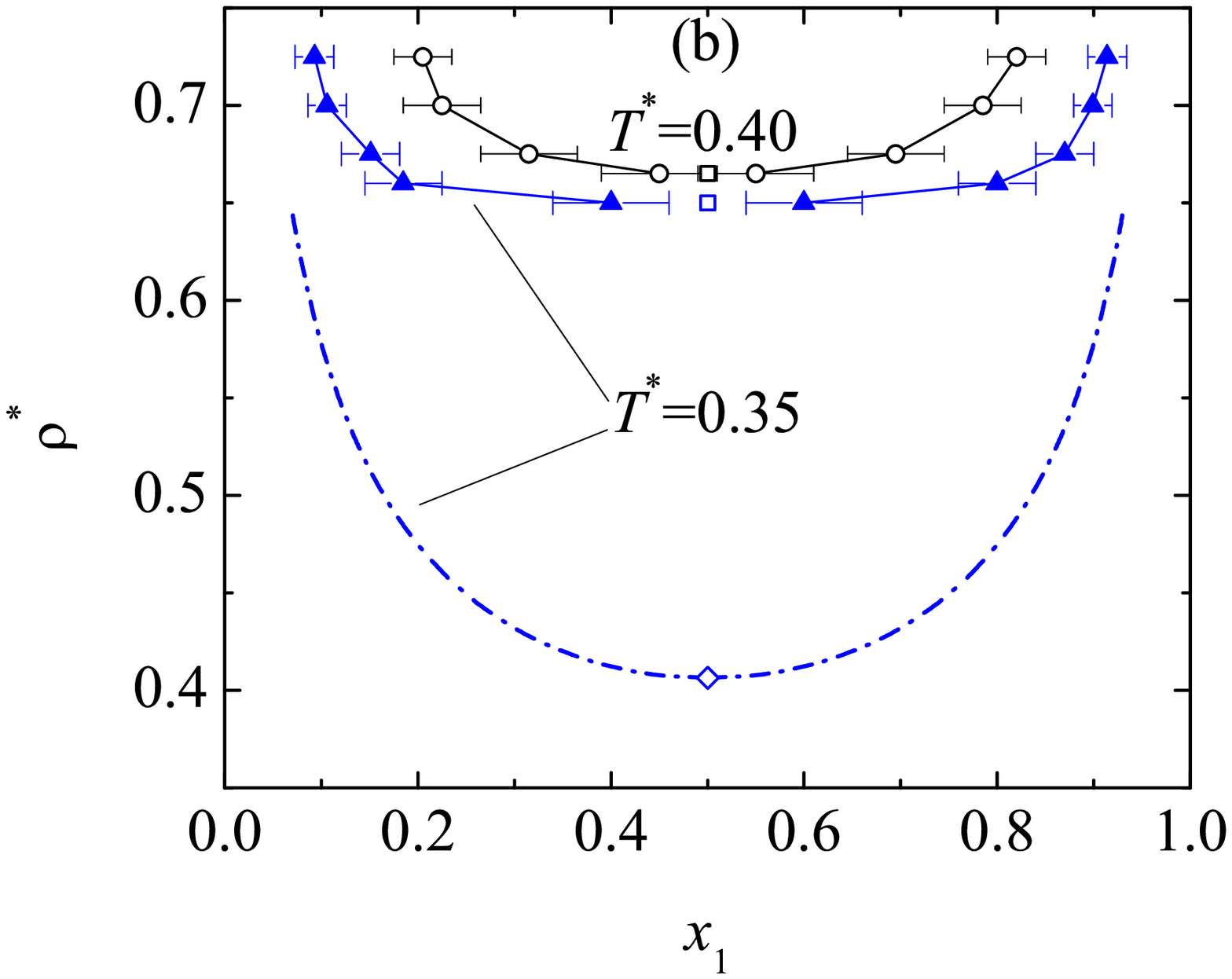}
\end{center}
\caption{Demixing curves for models (a) I0  and (b) B0  at two temperatures, as obtained from GEMC simulations,  in the density-mole fraction plane. In each case, the solid line is  a guide to the eye, {while the  critical  consolute point is represented by a square.} For model I0 we found $\rho_{cc}^*(T^*=0.4)=0.336$ and $\rho_{cc}^*(T^*=0.45)=0.429$; for model B0 the results are $\rho_{cc}^*(T^*=0.35)=0.650$  and $\rho_{cc}^*(T^*=0.4)=0.665$. {The dashed-dotted lines are the theoretical predictions (see Sec.\ \ref{theo_demix})}.}
\label{fig:fig4}
\end{figure}

\subsection{Demixing transition}
\label{sec:demixing}

\begin{table}
\caption{Demixing coexistence properties for models I0 and  B0, as obtained from our GEMC simulations. $T^*$ is the reduced temperature, $\rho^*$ is  the reduced density, and $x_1^{(\gamma)}$ is the mole fraction of species
$1$ in each one of the two coexisting phases $\gamma=$I, II. }
\label{tab:dem}
\begin{ruledtabular}
\begin{tabular}{cdddd}
Model&\multicolumn{1}{c}{$\;\;\;T^*$} &\multicolumn{1}{c}{$\;\;\;\rho^*$} & \multicolumn{1}{c}{$\;\;\;\;\;x_1^\text{(I)}$}&\multicolumn{1}{c}{$\;\;\;\;\;x_1^\text{(II)}$}\\
\hline
I0&0.4&0.7 & 0.005(5) & 0.992(5) \\
&&0.65 & 0.006(6) & 0.985(6) \\
&&0.6 & 0.01(1) & 0.97(1) \\
&&0.5 & 0.05(3) & 0.93(3) \\
&&0.4 & 0.19(4) & 0.81(4) \\
&&0.38 & 0.23(6) & 0.77(6) \\
&&0.36 & 0.32(9) & 0.68(9) \\
&&0.34 & 0.4(1) & 0.6(1) \\
&&&&\\
&0.45&0.7 & 0.01(1) & 0.99(1) \\
&&0.6 & 0.05(2) & 0.96(2) \\
&&0.5 & 0.14(4) & 0.87(4) \\
&&0.45 & 0.25(7) & 0.74(7) \\
&&0.43 & 0.4(1) & 0.6(1) \\
&&&&\\
B0&0.35&0.725 & 0.09(2) & 0.91(2) \\
&&0.7 & 0.11(2) & 0.90(2) \\
&&0.675 & 0.15(3) & 0.87(3) \\
&&0.66 & 0.18(4) & 0.80(4) \\
&&0.65 & 0.40(6) & 0.60(6) \\
&&&&\\
&0.4&0.725 & 0.20(3) & 0.82(3) \\
&&0.7 & 0.22(4) & 0.78(4) \\
&&0.675 & 0.31(5) & 0.69(5) \\
&&0.665 & 0.45(6) & 0.55(6) \\
\end{tabular}
\end{ruledtabular}
\end{table}

The bi-component nature of the systems raises the question of a possible demixing transition in which a rich-1 phase coexists with a rich-2 phase at a given temperature $T$, provided the density is larger than a certain  critical consolute density $\rho_{cc}(T)$.  The points $\rho_{cc}(T)$ or, reciprocally, $T_{cc}(\rho)$ define the so-called $\lambda$-line.\cite{W95} The interplay between the gas-liquid and demixing  transitions is a very interesting issue and  was discussed in a general framework by Wilding et al.\cite{WSN98}

Since all the spheres have the same size, a necessary condition for demixing {in the case of \emph{isotropic} potentials} is that the like attractions must be sufficiently stronger than the unlike attractions.\cite{WSN98,FGG05} {Assuming the validity of this condition to anisotropic potentials  and making a simple estimate based on the virial expansion, one finds} that demixing requires the coefficient of $x_1x_2$ in the second virial coefficient to be positive, i.e., $2e^{\epsilon_{11}/k_BT} >e^{\epsilon_{12}/k_BT}+e^{\epsilon_{21}/k_BT}$. While this demixing criterion is only approximate, it suggests that, out of the five models considered, only models B0 and I0 are expected to display demixing transitions.  {As a matter of fact, we have already discussed the spontaneous demixing phenomenon taking place in model I0 when a low-density phase and a high-density phase are in mutual equilibrium. In this section, however, we are interested in the segregation of the system, at a given  $T$ and for  $\rho>\rho_{cc}(T)$,  into a rich-2 phase I with $x_1^\text{(I)}=x_{d}(\rho)<\frac{1}{2}$ and a \emph{symmetric} rich-1 phase II with $x_1^\text{(II)}=1-x_{d}(\rho)>\frac{1}{2}$, both phases at the \emph{same} density.}

{Our GEMC simulation results are presented in Fig.\ \ref{fig:fig4} and Table \ref{tab:dem}.
We observe that,} as expected, $x_1^\text{(I)}=1-x_1^\text{(II)}$ within statistical fluctuations. {We have also checked that $\rho^\text{(I)}\simeq \rho^\text{(II)}$, {even though} this equality is not artificially enforced in the simulations}.
Such equality is also equivalent to $\rho^\text{(I)}\simeq\rho$ and we
checked that it was satisfied within a standard deviation
of $0.02\sigma^{-3}$ in all cases considered in Table \ref{tab:dem}.
To obtain the critical consolute density $\rho_{cc}^*$ for each temperature,  we extrapolated the data again according to the Ising scaling relation $\frac{1}{2}-x_d(\rho)=C(\rho-\rho_{cc})^{\beta_I}$.

It is interesting to note that just the absence of attraction when a particle of species 2 is below a particle of species 1 ($\epsilon_{21}=0$) in model B0 is sufficient to drive a demixing transition. However, as expected, at a common temperature (see $T^*=0.4$ in Fig.\ \ref{fig:fig4}), demixing requires higher densities in model B0 than in model I0.

{As said above, the interplay of condensation and demixing is an interesting problem by itself.\cite{WSN98,JF13} Three alternative scenarios are in principle possible for the intersection of the $\lambda$-line and the binodal curve: a critical end point, a triple point, or a tricritical point.\cite{WSN98} Elucidation of these scenarios would require grand canonical simulations (rather than GEMC simulations), what is beyond the scope of this paper.}

\section{Simple analytical theories\label{sec4}}

Let us now compare the above numerical results with simple theoretical predictions. The solution of integral equation theories for anisotropic interactions and/or multicomponent systems requires formidable numerical efforts, {with the absence of explicit expressions often hampering} physical insight. Here we want to deal with simple, purely analytical theories that yet include the basic ingredients of the models.

First, we take advantage of the short-range of the attractive well ($\Delta/\sigma=0.05$) to map the different SW interactions into  SHS interactions parameterized by  the ``stickiness'' parameters{\cite{MFGS13}}
\beq
t_{ij}\equiv \frac{1}{12\tau_{ij}}\equiv
\frac{\Delta}{\sigma}\left(1+\frac{\Delta}{\sigma}+\frac{\Delta^2}{3\sigma^2}\right)\left(e^{\epsilon_{ij}/k_BT}-1\right),
\label{tij}
\eeq
which combine the energy and length scales. This mapping preserves the exact second virial coefficient of the genuine SW systems, namely
\beq
\frac{B_2}{B_2^\hs}=1-3t_{11}+3x_1 x_2(2t_{11}-t_{12}-t_{21}),
\label{B2}
 \eeq
where $B_2^\hs=2\pi\sigma^3/3$ is the HS coefficient. The exact expression of the third virial coefficient $B_3$ in the SHS limit for arbitrary $t_{ij}$ is\cite{MFGS13}
\beqa
\frac{B_3}{B_3^\hs}&=&1-6t_{11}+\frac{72}{5} t_{11}^2-\frac{48}{5}t_{11}^3-\frac{6}{5}x_1x_2\Big[\left(12t_{11}-{5}\right)\nn
&&\times\left(2t_{11}-t_{12}-t_{21}\right)-8t_{11}\left(t_{11}^2-t_{12}t_{21}\right)\nn
&&-2(4t_{11}-3)\left(2t_{11}^2-t_{12}^2-t_{21}^2\right)+2\alpha\left(t_{12}-t_{21}\right)^2\Big],\nn
\label{2}
\eeqa
where $B_3^\hs=5{\pi^2\sigma^6}/{18}$ and
\beq
\alpha\equiv \frac{3\sqrt{3}}{\pi}-1.
\label{alpha}
\eeq

\subsection{Equations of state}

One advantage of the $\text{SW}\to\text{SHS}$ mapping is that the Percus--Yevick (PY) integral equation is exactly solvable for SHS mixtures with \emph{isotropic} interactions ($t_{12}=t_{21}$).\cite{PS75,B75} In principle, that solution can be applied to the models SW, I0, and J0 represented in Fig.\ \ref{fig:fig2}. On the other hand, if $t_{11}\neq 0$ (models SW and I0), the PY solutions are related to algebraic equations of second (SW) or fourth (I0) degrees, what creates the problem of disappearance
of the physical solution for large enough densities or stickiness. In particular, we have observed that the breakdown of the solution preempts the existence of a critical point in model I0. However, in the case of model J0 ($t_{11}=0$, $t_{12}=t_{21}=t$), the PY solution reduces to a \emph{linear} equation whose solution is straightforward. Following the virial {($v$)} and the energy {($u$)} routes, the respective expressions for the compressibility factor $Z\equiv P/\rho k_BT$ (where $P$ is the pressure) have the form
\beq
Z_v(\eta,t,x_1)=Z_v^\hs(\eta)-x_1 x_2 Z_v^{(1)}(\eta,t)-x_1^2 x_2^2 Z_v^{(2)}(\eta,t),
\label{1}
\eeq
\beq
Z_u(\eta,t,x_1)=Z_u^\hs(\eta)-x_1 x_2 Z_u^{(1)}(\eta,t),
\label{4}
\eeq
where  $\eta=\pi\rho^*/6$ is the packing fraction,
\beq
Z_v^\hs(\eta)=\frac{1+2\eta+3\eta^2}{(1-\eta)^2}
\eeq
is the HS compressibility factor derived from the PY equation via the virial route,
$Z_u^\hs$ is an indeterminate integration constant, and
the explicit expressions for $Z_v^{(1)}$, $Z_v^{(2)}$, and $Z_u^{(1)}$ are
\beqa
Z_v^{(1)}(\eta,t)&=&\frac{24\eta t}{(1-\eta+6\eta t)^2}\left[\frac{1+2\eta}{1-\eta}+3\eta t \frac{2+2\eta-5\eta^2/2}{(1-\eta)^2}\right.\nn
&&\left.+6\eta^2 t^2 \frac{2-4\eta-7\eta^2}{(1-\eta)^3}\right],
\label{2b}
\eeqa
\beqa
Z_v^{(2)}(\eta,t)&=&\frac{288\eta^3 t^2(2+\eta)}{(1-\eta+6\eta t)^3}\left[\frac{1}{1-\eta}- t \frac{2-11\eta}{(1-\eta)^2}\right.\nn
&&\left.+ t^2 \frac{2-10\eta+61\eta^2/2}{(1-\eta)^3}\right],
\label{3b}
\eeqa
\beq
Z_u^{(1)}(\eta,t)=\frac{6\eta}{(1-\eta)^2}\left[\frac{2t(2+\eta)}{1-\eta+6\eta t}+\ln\frac{1-\eta+6\eta t}{1-\eta}\right].
\label{5b}
\eeq
To the best of our knowledge, this extremely simple solution of the PY integral equation for a model of SHS mixtures had not been unveiled before.

{As apparent from Fig.\ \ref{fig:fig2}, model A0 is  a close relative} of model J0. However, the fact that $\epsilon_{12}\neq\epsilon_{21}=0$ (or $t_{12}\neq t_{21}=0$) makes the interaction anisotropic and prevents the PY equation from being exactly solvable in this case. On the other hand, we have recently proposed\cite{MFGS13} a simple rational-function approximation (RFA)  that applies to models with $t_{12}\neq t_{21}$ and reduces to the PY solution in the case of isotropic models ($t_{12}=t_{21}$). The RFA solution for model A0 yields once more  a linear equation. The virial and energy equations of state are again of the forms \eqref{1} and \eqref{4}, respectively, with expressions for $Z^{(1)}_v$, $Z^{(2)}_v$, and   $Z^{(1)}_u$ given by
\beq
Z_v^{(1)}(\eta,t)=\frac{12\eta t}{1-\eta+6\eta t}\left[\frac{1+2\eta}{(1-\eta)^2}+2\eta t \frac{1-2\eta-7\eta^2/2}{(1-\eta)^3}\right],
\label{3}
\eeq
\beq
Z_v^{(2)}(\eta,t)=\frac{72\eta^3 t^2(2+\eta)}{(1-\eta)^3(1-\eta+6\eta t)},
\label{4b}
\eeq
\beq
Z_u^{(1)}(\eta,t)=\frac{3\eta}{(1-\eta)^2}\left[\frac{2t(2+\eta)}{1-\eta+6\eta t}+\ln\frac{1-\eta+6\eta t}{1-\eta}\right].
\label{5}
\eeq
In the RFA solution for model A0 the exact third virial coefficient \eqref{2} is recovered by the interpolation formula \beqa
Z&=&Z_\cs^\hs+\alpha\left(Z_v-Z_v^\hs\right)+(1-\alpha)\left(Z_u-Z_u^\hs\right)\nn
&=&Z_\cs^\hs-x_1 x_2\left[\alpha Z_v^{(1)}+(1-\alpha)Z_u^{(1)}\right]-x_1^2x_2^2\alpha Z_v^{(2)},\nn
\label{6}
\eeqa
where
\beq
Z_\cs^\hs(\eta)=\frac{1+\eta+\eta^2-\eta^3}{(1-\eta)^3}
 \eeq
is the HS Carnahan--Starling compressibility factor  and the interpolation weight $\alpha$ is given by Eq.\ \eqref{alpha}. By consistency, Eq.\ \eqref{6} will also be employed  in the PY solution of model J0.

In the cases of models with $\epsilon_{11}\neq 0$ (i.e., SW, B0, and I0), the PY and RFA theories fail to have physical solutions in  regions of the temperature-density plane overlapping with the gas-liquid transition. In order to circumvent this problem, we adopt here a simple perturbative approach:
\beq
Z=Z^\text{ref}+\left(B_2-B_2^\text{ref}\right)\rho+\left(B_3-B_3^\text{ref}\right)\rho^2,
\eeq
where $Z^\text{ref}$ is the compressibility factor of a reference model and $B_2^\text{ref}$ and and $B_3^\text{ref}$ are the associated virial coefficients. As a natural choice (see {Fig.\ \ref{fig:fig2}}), we take the models J0, A0, and HS {({which lie} on the plane $\epsilon_{11}/\epsilon=0$)} as reference systems for the models SW, B0, and I0 {({which lie} on the plane $\epsilon_{11}/\epsilon=1$)}, respectively. More specifically,
\beq
Z^{\text{SW}}=Z^{\text{J0}}+\left(B_2^{\text{SW}}-B_2^{\text{J0}}\right)\rho+\left(B_3^{\text{SW}}-
B_3^{\text{J0}}\right)\rho^2,
\label{7}
\eeq
\beq
Z^{\text{B0}}=Z^{\text{A0}}+\left(B_2^{\text{B0}}-B_2^{\text{A0}}\right)\rho+\left(B_3^{\text{B0}}-
B_3^{\text{A0}}\right)\rho^2,
\label{8}
\eeq
\beq
Z^{\text{I0}}=Z_\cs^{\text{HS}}+\left(B_2^{\text{I0}}-B_2^{\text{HS}}\right)\rho+\left(B_3^{\text{I0}}-
B_3^{\text{HS}}\right)\rho^2.
\label{9}
\eeq
Here, $Z^{\text{J0}}$ and $Z^{\text{A0}}$ are given by Eq.\ \eqref{6} (with the corresponding expressions of $Z_v^{(1)}$, $Z_v^{(2)}$, and $Z_u^{(1)}$) and the virial coefficients are obtained in each case from Eqs.\ \eqref{B2} and \eqref{2} with the appropriate values of $t_{11}$, $t_{12}$, and $t_{21}$.

{}From the explicit knowledge of $Z(\eta,t,x_1)$, standard thermodynamic relations allow one to obtain the free energy per particle $a(\eta,t,x_1)$ and the chemical potentials $\mu_i(\eta,t,x_1)$ as
\beqa
\beta a(\eta,t,x_1)&=&\int_0^\eta d\eta'\,\frac{Z(\eta',t,x_1)-1}{\eta'}+x_1\ln(x_1\eta)\nn
&&+(1-x_1)\ln[(1-x_1)\eta]+\text{const},
\label{10}
\eeqa
\beqa
\beta \mu_1(\eta,t,x_1)&=&\beta a(\eta,t,x_1)+Z(\eta,t,x_1)\nn
&&+(1-x_1)\frac{\partial \beta a(\eta,t,x_1)}{\partial x_1},
\label{11}
\eeqa
\beq
\mu_2(\eta,t,x_1)=\mu_1(\eta,t,1-x_1),
\label{12}
\eeq
where $\beta\equiv 1/k_BT$.

\subsection{Gas-liquid coexistence}

The critical point $(\eta_c,t_c)$ of the gas-liquid transition is obtained from the {well-known condition that the critical isotherm in the pressure-density plane presents an inflection point with horizontal slope at the critical density.\cite{HM06} In terms of the compressibility factor $Z$, this implies}
\beq
\left.\frac{\partial \left[\eta Z(\eta,t_c,1/2)\right]}{\partial \eta}\right|_{\eta=\eta_c}=\left.\frac{\partial^2 \left[\eta Z(\eta,t_c,1/2)\right]}{\partial \eta^2}\right|_{\eta=\eta_c}=0,
\label{13}
\eeq
where  equimolarity ($x_1=\frac{1}{2}$) has been assumed.
For temperatures below the critical temperature (i.e., $t>t_c$) the packing fractions $\eta_g$ and $\eta_l$ of the gas and liquid coexisting phases are obtained from the conditions {of equal pressure (mechanical equilibrium) and equal chemical potential (chemical equilibrium),\cite{HM06} i.e.}
\beq
\eta_g Z(\eta_g,t,1/2)=\eta_l Z(\eta_l,t,1/2),
\label{14.1}
\eeq
\beq
\mu_1(\eta_g,t,1/2)=\mu_1(\eta_l,t,1/2).
\label{14}
\eeq

\begin{table}
\caption{Comparison between the critical points measured in simulations with those obtained from theoretical approaches. }
\label{tab:crit}
\begin{ruledtabular}
\begin{tabular}{c c cccc}
Method& SW&B0&I0&J0&A0\\
\hline
\multicolumn{6}{c}{$T_c^*$}\\
Simulation&$0.369$\footnotemark[1]&$0.338$\footnotemark[2]&$0.368$\footnotemark[2]&$0.258$\footnotemark[2]&$0.193$\footnotemark[2]\\
Our theory& $0.377$ &$0.341$& $0.331$& $0.278$ & $0.214$\\
Noro--Frenkel    & $0.369$& $0.335$& $0.297$& $0.297$& $0.247$\\
\multicolumn{6}{c}{$\rho_c^*$}\\
Simulation&$0.508$\footnotemark[1]&$0.373$\footnotemark[2]&$0.344$\footnotemark[2]&$0.344$\footnotemark[2]&$0.342$\footnotemark[2]\\
Our theory& $0.356$ &$0.330$& $0.366$& $0.376$ & $0.359$\\
\end{tabular}
\end{ruledtabular}
\footnotetext[1]{GCMC results for the one-component SHS fluid From Ref.\ \protect\onlinecite{MF04}}
\footnotetext[2]{Our GEMC simulation results}
\end{table}

In order to make contact with the GEMC results, the theoretical values of $t_c$ have been mapped onto those of $T_c^*$ by inverting Eq.\ \eqref{tij}, namely
\beq
\frac{1}{T^*}=\ln\left[1+\frac{t}{({\Delta}/{\sigma})\left(1+{\Delta}/{\sigma}+{\Delta^2}/{3\sigma^2}\right)}\right]
\eeq
with $\Delta/\sigma=0.05$.

\begin{figure}
  \includegraphics[width=8cm]{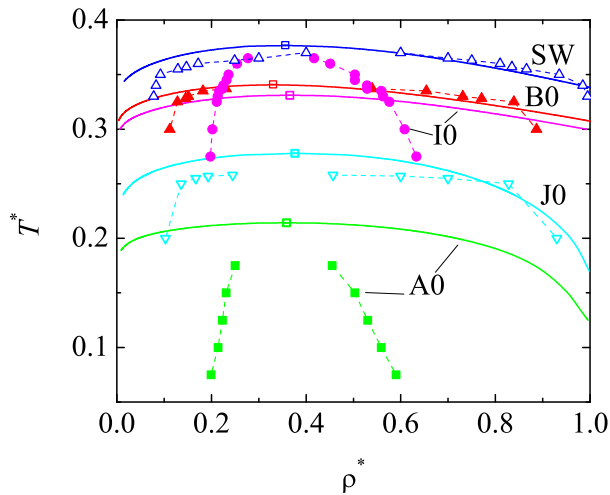}
\caption{Gas-liquid binodals for  models SW, A0, B0, I0, and J0, as obtained from our theoretical method {(solid lines)}. The critical points are represented by  {open} squares. {The symbols joined by dashed lines correspond to our GEMC data (see Fig.\ \ref{fig:fig3}).}}
\label{fig:fig5}
\end{figure}

Table \ref{tab:crit} compares the critical points obtained in simulations for the one-component SW fluid (in the SHS limit) and for models B0, I0,  J0, and A0  (see Fig.\ \ref{fig:fig2}) with those stemming from our simple theoretical method. Results from the Noro--Frenkel (NF) corresponding-state criterion,\cite{NF00} according to which $B_2/B_2^\hs=-1.21$ at the critical temperature, are also included.
We observe that, despite its simplicity and the lack of fitting parameters, our fully analytical theory predicts quite well the location of the critical point, especially in the case of $T_c^*$. It improves the estimates obtained from the NF criterion, except in the SW case, where, by construction, the NF rule gives the correct value.
In what concerns the gas-liquid binodals,  Fig.\ \ref{fig:fig5} shows that the theoretical curves agree fairly well with the GEMC data, except in the cases of models I0 and A0, where the theoretical curves are  much flatter than the simulation ones. {The lack of agreement with the binodal curve of model I0 can be partially due to the fact that in the theoretical treatment the two coexisting phases are supposed to be equimolar, while this is not the case in the actual simulations (see Table \ref{table2b}).}

\subsection{Demixing transition}
\label{theo_demix}

In the case of the demixing transition, the critical consolute density $\eta_{cc}$ at a given temperature is obtained from
\beq
\left.\frac{\partial^2 a(\eta_{cc},t,x_1)}{\partial x_1^2}\right|_{x_1=\frac{1}{2}}=0.
\label{15}
\eeq
For $\eta>\eta_{cc}$, the demixing mole fraction $x_1=x_d(\eta)$ is the solution to
\beq
\mu_1(\eta,t,x_d)=\mu_1(\eta,t,1-x_d).
\label{16}
\eeq
In terms of the compressibility factor $Z$, Eqs.\ \eqref{15} and \eqref{16} can be rewritten as
\beq
\int_0^{\eta_{cc}} d\eta\,\frac{\left.\partial^2 Z(\eta,t,x_1)/\partial x_1^2\right|_{x_1=\frac{1}{2}}}{\eta}=-4,
\label{17}
\eeq
\beq
\int_0^\eta d\eta'\,\frac{\partial Z(\eta',t,x_d)/\partial x_d}{\eta'}=\ln\frac{1-x_d}{x_d},
\label{18}
\eeq
respectively.

The perturbative approximations for models I0 and B0 succeed in predicting demixing transitions, even though their respective reference systems (HS and A0) do not demix. In the case of model I0, the  critical consolute densities are $\rho_{cc}^*(T^*=0.4)=0.306$ and  $\rho_{cc}^*(T^*=0.45)=0.390$,  which are about 9\% lower than the values obtained in our GEMC simulations. In the case of model B0, our simple theory predicts a critical consolute point only if $t>0.7667$, i.e., if $T^*<0.364$, so no demixing is predicted at $T^*=0.4$, in contrast to the results of the simulations. At $T^*=0.35$ the theoretical prediction is $\rho_{cc}^*=0.406$, a value about 39\% smaller than the GEMC one.
{The theoretical demixing curves at $T^*=0.4$ and $T^*=0.45$ for model I0 and at $T^*=0.35$ for model B0 are compared with the GEMC results in Fig.\ \ref{fig:fig4}. We can observe a fairly good agreement in the case of model I0, but not for model B0.} {In the latter case, the theoretical curve spans a density range comparable to that of model I0, while simulations show a much flatter demixing curve.}

\section{Concluding remarks\label{sec5}}
In conclusion, we have proposed a novel class of binary-mixture Janus fluids with up-down  constrained orientations. The class encompasses, as particular cases, the conventional one-component SW fluid,  mixtures  with isotropic attractive interactions only between like particles (model I0) or unlike particles (model J0), and genuine Janus fluids with anisotropic interactions and different patch-patch affinities (models A0 and B0).  Both GEMC numerical
simulations and simple theoretical approximations  have been employed to analyze the gas-liquid transition under {\emph{global}} equimolar conditions for the five models and the demixing transition for the two models (I0 and B0) where the attraction between like particles is stronger than between unlike ones.
The theoretical analysis employed a mapping onto SHS interactions, that were then studied by means of the PY theory (model J0), the RFA (model A0), and low-density virial corrections (models SW, I0, and B0), with
semi-quantitative agreement with numerical simulations.

Interestingly, the presence of attraction in only one out of the four possible patch-patch interactions (model A0) turns out to be enough to make the gas-liquid transition possible.
Reciprocally, the lack of attraction in only one of the two possible patch-patch interactions between unlike particles (model B0) is enough to produce a demixing transition.
{Except in model I0, the coexisting gas and liquid phases have an equimolar composition.}
As the average attraction is gradually  decreased,  the gas-liquid critical point shifts to lower
temperatures (except for an interesting inversion of tendency observed when going
from the isotropic model I0  to the anisotropic model B0) and lower densities. Moreover, the coexistence region progressively shrinks, in analogy with what is observed in the unconstrained
one-component  Janus fluid\cite{FDGMP07,GRSG12}  and in the empty liquid scenario.\cite{BLZS06}
On the other hand, the imposed constraint in the
orientation of the attractive patches does not allow for the formation
of those inert clusters\cite{SGP09,FGSP11,F12} which in the original Janus
fluid are responsible for a re-entrant gas branch.\cite{SGP09,SGP10,RWDCVL11}

\acknowledgments
{The authors are grateful to J.-P. Hansen for useful suggestions.} R.F. acknowledges the use of the PLX computational facility of CINECA through the ISCRA call. A.G. acknowledges funding from  PRIN-COFIN2010-2011 (contract 2010LKE4CC).
The research of M.A.G.M. and A.S. has been supported by the Spanish government  through Grant No.\ FIS2010-16587 and by the Junta de Extremadura (Spain) through Grant No.\ GR101583, partially financed by FEDER funds. M.A.G.M is also grateful to the Junta de Extremadura (Spain) for  the pre-doctoral fellowship PD1010.




\end{document}